\documentclass[preprint,12pt]{elsarticle}
\usepackage{etoolbox}
\makeatletter
\def\ps@pprintTitle{%
	\let\@oddhead\@empty
	\let\@evenhead\@empty
	\def\@oddfoot{\reset@font\hfil\thepage\hfil}
	\let\@evenfoot\@oddfoot
}
\makeatother
\usepackage{graphics}
\usepackage[table,xcdraw]{xcolor}
\usepackage{amsmath, amsthm, amssymb}
\usepackage{natbib}
\usepackage[colorlinks=true]{hyperref}
\usepackage{lscape}
\usepackage[utf8]{inputenc} 
\usepackage{amsthm}

\usepackage{enumerate}
\usepackage{mathrsfs}
\usepackage{setspace}
\usepackage{multirow}
\usepackage{graphicx}
\usepackage{amsfonts}
\usepackage{verbatim}
\usepackage{amssymb}
\usepackage{amsthm}
\usepackage{multirow,bigstrut,bigdelim}

\theoremstyle{plain}

\theoremstyle{remark}

\numberwithin{equation}{section}
\usepackage[mathscr]{euscript}
\usepackage{geometry} 
\usepackage{graphicx,lscape} 
\usepackage{booktabs}
\usepackage{array}
\usepackage{paralist} 
\usepackage{verbatim}
\usepackage{subfig} 
\biboptions{comma,round,authoryear}

\begin{document}
	\begin{frontmatter}
		
		\title{\textbf{A quantile-based bivariate distribution}}		\author{Shifna. P. R.\corref{cor1}}
		\ead{shifnarazack92@gmail.com}
		\author{N. Unnikrishnan Nair}
		\ead{unnikrishnannair4@gmail.com}
		\author{S. M. Sunoj}
		\ead{smsunoj@cusat.ac.in}
		
		\cortext[cor1]{Corresponding author}
		\address{Department of Statistics\\Cochin University of Science and Technology\\Cochin 682 022, Kerala, India.}

		\begin{abstract}
			In this paper we present a flexible bivariate distribution specified by a quantile function. The distribution contains as special cases new bivariate exponential, Pareto I, Pareto II, beta, power, log logistic and uniform distributions and also can approximate many other continuous models. Various $L$-moment based properties of the distribution such as covariance, coskewness, cokurtosis, $L$-correlation, etc are discussed. The distribution is used to model two real data sets.
		\end{abstract}
		\begin{keyword}
			Quantile function\sep $L$-moments \sep $L$-correlation\sep Bivariate distributions.
			\MSC[2020] 62E05
		\end{keyword}	
	\end{frontmatter}
	\section{Introduction}\label{sec 1}
	Quantile functions play the same role as distribution functions, they being equivalent representations of one another and so quantile functions have applications in all problems where distribution functions are useful. Although generally a quantile function is defined as the inverse of the distribution function and obtained in this way on many occasions, several distributions are defined in terms of quantile functions independently of the form of the distribution function as a left continuous, non-decreasing function on the unit interval, $I=[0,1]$. Examples of such distributions are the lambda distributions and its variants, Wakeby, power-Pareto, Govindarajulu, Kamps, linear mean residual quantile function models etc. See \citet{nair2022modelling} for details. The popularity of such quantile forms arises in model building from the fact that they provide satisfactory approximations for a wide variety of data, have simple algebraic structure, easy to use in generating random samples in simulation studies and often estimates based on them are more robust than their counterparts derived from distribution functions. Moreover simple algebraic operations like addition, multiplication and monotone transformations on quantile functions lead to more general models which are quiet handy in data analysis problems.
	
	There is an alternative route by which quantile-based distributions can be visited by writing the density function in terms of the distribution function. A general discussion on the class of distributions defined by the relationship between the density and distribution functions initiated in \citet{jones2007class} points out the members of this class and their properties. It is pointed out that the distribution of order statistics belongs to this class, see \citet{jones2004families} and the discussions on this paper. Special properties of the distribution of order statistics of some specified distributions can also be seen in \citet{kamps1991general}, \citet{balakrishnan2003characterization} and \citet{akhundov2004new}.
	
	Inspite of the utility of quantile functions in analyzing univariate data is well established, the progress of similar approaches in the multivariate case has been slow. Some attempts made by researchers like \citet{chen2002distribution}, \citet{belzunce2007quantile} and \citet{cai2010multivariate} to extend the concept of quantiles to higher dimensions, were not directed towards suggesting multivariate quantile functions that can specify distributions as in the univariate case. Conceiving the bivariate quantile function as a pair that transforms the points in the unit square $I^2$ to points in the $x_1-x_2$ plane, in the same way as I is transformed to $\mathbb{R}$ in the univariate cases, \citet{vineshkumar2019bivariate} and \citet{unnikrishnan2023properties} find bivariate quantile function of distributions in $\mathbb{R}_2$. They have pointed out the basic properties of these functions and their applications as model of bivariate lifetime data. In the process some bivariate distributions generated by quantile functions were also presented.
	
	The main objective of the present work is to propose a flexible bivariate distribution arising from a bivariate quantile function, that can represent different types of data in a unified framework. The proposed distribution subsumes some new bivariate exponential, Pareto, power, log logistic, etc as special cases and many other quantile functions with special properties as other members. With its flexibility it can serve as a black box model of random phenomena in a wide range of cases. A method of inferring the parameters with real data illustrations also form part of this work.
	
	The work is organized into five sections. In the next section we introduce the model and deduce its special cases . This is followed in Section \ref{sec 3} by discussing the properties of the model including covariance, coskewness, cokurtosis and measures of dependence. The application of the model in two real data situations is demonstrated in Section \ref{sec 4}. The paper ends with  Section \ref{sec 5} in the form of a short conclusion.

	\section{The bivariate distribution}\label{sec 2}
	We consider a random vector $\left(X_1, X_2\right)$ with absolutely continuous distribution function, $F\left(x_1, x_2\right)$ and survival function $\bar F\left(x_1, x_2\right)$. Then the bivariate quantile function of $\left(X_1, X_2\right)$
	is defined with respect to the marginal distribution function $F_1\left(x_1\right)$ of $X_1$ and the conditional distribution function $F_{21}\left(x_2 \mid x_1\right)=P\left(X_2 \leq x_1 \mid X_1>x_1\right)$ as the pair $\left(Q_1\left(u_1\right), Q_{21}\left(u_1, u_2\right)\right)$ where 
	\[
	Q_1\left(u_1\right)=\inf \left[x_1\mid F\left(x_1\right) \geq u_1\right],\quad 0 \leq u_1 \leq1
	\]
	and
	\[
	Q_{21}\left(u_1, u_2\right)=\inf\left[x_2 \mid F_{21}\left(x_2 \mid Q_1\right) \geq u_2\right],\quad 0 \leq u_2 \leq 1.
	\]
	The bivariate survival function of $\left(X_1, X_2\right)$ is recovered from $\left(Q_1, Q_{21}\right)$ as
	$$
	\bar{F}\left(x_1, x_2\right)=\bar F_1\left(x_1\right) \bar{F}_{21}\left(x_2 \mid x_1\right)=P\left(X_1 \geq x_1\right) P\left(X_2>x_1 \mid X_1>x_1\right),
	$$
	where $F_1\left(x_1\right)=\inf \left[u_1 \mid Q_1\left(u_1\right) \geq x_1\right]$ and $F_{21}\left(x_2 \mid x_1\right)=\inf \left[u_2 \mid Q_{21}\left(u_1, u_2\right)\geq x_2\right]$.
	Further the quantile density functions $q_1\left(u_1\right)=\frac{d Q_1}{d u_1}$ and
	$q_{21}\left(u_1, u_2\right)=\frac{\partial Q_{21}}{\partial u_2}$ also determine the distribution of $\left(X_1, X_2\right)$ uniquely.
	
	The bivariate distribution defined in this
	work is characterized by the pair $\left(q_1(u_1), q_{21}(u_1,u_2)\right)$ where
	\begin{equation}\label{2.1}
		q_1\left(u_1\right)=c_1 u_1^{\alpha_1}\left(1-u_1\right)^{\beta_1}
	\end{equation}
	and 
	\begin{equation}\label{2.2}
		q_{21}\left(u_1, u_2\right)=c_2\left(1+\theta u_1\right) u_2^{\alpha_2}\left(1-u_2\right)^{\beta_2},
	\end{equation}
	where $c_1, c_2>0,\ \theta \geq 0$ and $\alpha_i$ and $\beta_i$ are real parameters, $i=1,2$.
	As $u_1 \rightarrow 0$ in $q_{21}$, $q_2\left(u_2\right)=c_2 u_2^{\alpha_2}\left(1-u_2\right)^{\beta_2}$ represents the marginal distribution of $X_2$.
	
	An attractive aspect of the model in (\ref{2.1}) and (\ref{2.2}) is that it gives many tractable distributions as special cases, that have different shapes and properties that makes it a highly flexible family.
	
	\textbf{Case (i):} $\alpha_i, \beta_i>-1$.
	
	In this case we have
	\begin{equation}\label{2.3}
		Q_i\left(u_i\right)=C_i B_{u_i}\left(\alpha_i+1, \beta_i+1\right),\quad i=1,2  
	\end{equation}
	and
	\begin{equation}\label{2.4}
		Q_{21}\left(u_1, u_2\right)=C_2\left(1+\theta u_1\right) B_{u_2} \left(\alpha_2+1, \beta_2+1\right),
	\end{equation}
	where  $B_x(m, n)=\int_0^x t^{m-1}(1-t)^{n-1}dt$ is the incomplete beta function. Note that $C_i=\frac{1}{B\left(\alpha_i+1, \beta_i+1\right)}$,  so that (\ref{2.4}) is the incomplete beta function ratio $I_u\left(\alpha_i+1, \beta_i+1\right)=$ $\frac{B_u\left(\alpha_i+1,\beta_i+1\right)}{B\left(\alpha_i+1, \beta_i+1\right)}.$ Hence the distribution function of $X_i$ is the inverse, $I^{-1}_{u_i}\left(\alpha_i+1, \beta_i+1\right)$, of the incomplete beta function ratio with support $[0,1]$. This distribution called the complementary beta distribution has been studied in \citet{jones2002complementary}. For purposes of computation, $F_{X_i}\left(u_i\right)=I^{-1}_ {u_i}\left(\alpha_i+1, \beta_i+1\right)$ is available in tabular form in packages as inv inc beta $(a, b)$. The conditional distribution $F_{21}$ is also of the same form with
	\begin{equation*}
		F_{X_2\mid Q_1}\left(x_2\right)=x_2 \frac{I^{-1}_ {u_2}\left(\alpha_2+1, \beta_2+1\right)}{\left(1+\theta u_1\right)}.
	\end{equation*}
	We name the bivariate distributions in this paper according to the form of the marginals and accordingly when $\alpha_i>-1,\ \beta_i>-1$ we have a bivariate complementary beta distribution with support $I^2$.
	
	\textbf{Case (ii):} $\beta_i=0$.
	
	We have $q_1\left(u_1\right)=c_1 u_1^{\alpha_1} ;\quad  q_{21}=c_2\left(1+\theta u_1\right) u_2^{\alpha_2},$
	\[
	\Rightarrow Q_1\left(u_1\right)=\frac{c_1 u_1 ^{\alpha_1+1}}{\alpha_1+1}; \quad Q_{21}\left(u_1, u_2\right)=\frac{c_2\left(1+\theta u_1\right) u_2^{\alpha_2+1}}{\alpha_2+1}, 
	\]
	\[
	\Rightarrow F_1\left(x_1\right)=\left(\frac{x_1}{b_1}\right)^{a_1} ; \quad F_{21}\left(x_2 \mid x_1\right)=\left(\frac{x_2}{\left(1+\theta u_1\right)b_2}\right)^{a_2},
	\]
	where $ a_i=\left(\alpha_i+1\right)^{-1}>0 ; \quad b_i=\frac{c_i}{\alpha_1+1}>0 ; \quad u_1=F_1\left(x_1\right) $.
	\begin{equation}\label{2.5}
			\begin{aligned}
				\Rightarrow \bar{F}\left(x_1, x_2\right)=\left(1-\left(\frac{x_1}{b_1}\right)^{a_1}\right)&\left(1-\left(1+\theta\left(\frac{x_1}{b_1}\right)^{a_1}\right)^{-a_2}\left(\frac{x_2}{b_2}\right)^{a_2}\right), \\
				& 0 \leq x_i \leq b_i;\quad a_i, b_i>0.
			\end{aligned}   
	\end{equation}
	Equation (\ref{2.5}) represents a bivariate power distribution with marginals $F_i\left(x_i\right)=\left(\frac{x_i}{b_i}\right)^{a_i},$ $0 \leq x_i \leq b_i;\ a_i>0;\ b_i>0$.
	When $a_i=1$, that is $\alpha_i=0$ the special case of a bivariate uniform distribution $\bar{F}\left(x_1, x_2\right)=\left(1-\frac{x_1}{b_1}\right)\left(1-\frac{x_2}{b_2\left(1+\theta\frac{x_1}{ a_1}\right)} \right)$ is obtained. In the above case $b_i>0 \Rightarrow \alpha_i>-1$. But distributions can be defined for $\alpha_i \leq-1$ in which case the supports become negative for $X_1$ and $X_2$.
	
	\textbf{Case (iii):} $\alpha_i=0,\ \beta_i=-1$.
	
	Calculations similar to the above case leads to the bivariate exponential distribution
	\begin{equation}\label{2.6}
		\bar{F}\left(x_1, x_2\right)=\exp \left[-\frac{x_1}{c_1}-\frac{x_2}{c_2\left(1+\theta\left(1-e^{-x_1} \right)\right.}\right].   
	\end{equation}
	The marginals are exponential with $\bar{F}_i\left(x_i\right)=\exp \left[-\frac{x_i}{c_i}\right]$. The conditional distribution of $X_2$ given $X_1 > x_1$ is
	$$
	\bar F_{21}\left(x_2 \mid x_1\right) = \exp \left[\frac{-x_2}{c_2(1+\theta u_1)}\right],\quad  u_1=1-e^{-\frac{x_1}{c_1}}.
	$$
	
	\textbf{Case (iv):} $\alpha_i=0,\ \beta_i>-1$
	
	With $$q_1\left(u_1\right)=c_1\left(1-u_1\right)^{\beta_1}\ \text{and} \ q_{21}\left(u_1, u_2\right)=c_2\left(1+\theta u_1\right)\left(1-u_2\right)^{\beta_2},$$
	we get
	$$
	\begin{aligned}
		&u_1= \bar F_1(x_1)=\left(1-\frac{x_1}{b_1}\right)^{a_1}, \quad b_1=(\frac{1+\beta_1}{c_1})^{-1}\ge 0,\quad a_1=\left(1+\beta_1\right)^{-1},\\
		&\bar F_{21}(x_2\mid x_1)=\left(1-\frac{x_2}{b_2(1+\theta u_1)}\right)^{a_2}, \quad b_2=(\frac{1+\beta_2}{c_2})^{-1},\quad a_2=\left(1+\beta_2\right)^{-1}.
	\end{aligned}
	$$
	Thus
	\begin{equation}\label{2.7}
		\begin{aligned}
			\bar F(x_1,x_2)=\left(1-\frac{x_1}{b_1}\right)^{a_1} &\left(1-\frac{x_2}{b_2(1+\theta(1-(1-\frac{x_1}{b_1})^{a_1})}\right)^{a_2}, \; a_1,\ b_1,\ a_2,\ b_2>0,\\
			&0 \leq x_1 \leq b_1 ;\quad 0 \leq x_2 \leq b_2\left(1+\theta\left(1-\left(1-x_1\right)^{b_1}\right)\right. .
		\end{aligned}  
	\end{equation}
	We have (\ref{2.7}) as a bivariate rescaled beta distribution.
	
	\textbf{Case (v):} $\alpha_i=0, \beta_i<-1<0$.
	
	The calculations are very similar to case (iv).
	\begin{equation}\label{2.8}
		\bar{F}\left(x_1, x_2\right)=\left(1+\frac{x_1}{b_1}\right)^{-d_1}\left(1+\frac{x_2}{b_2(1+\theta(1-(1+\frac{x_1}{b_1})^{-d_1}))^{-d_2}}\right), \quad x_1, x_2>0.    
	\end{equation}
	with $d_i=-(1+\beta_i)^{-1}> 0$ and $b_i=\frac{c_i}{d_i}> 0$. It is easy to
	see that (\ref{2.8}) is a bivariate Pareto II (Lomax) law.
	
	\textbf{Case (vi):}  $\alpha_i=0, \ \beta_i>-1$
	
	Reparametrizing $c_i=\frac{\sigma_i}{\alpha_i}$ and $\beta_i=\frac{1}{\alpha_ i}-1$,
	$$
	q_1\left(u_1\right)=\frac{\sigma_1}{\alpha_1}\left(1-u_1\right)^{-\frac{1}{\alpha_1}-1},
	$$
	which is the quantile density function of the Pareto I distribution with survival function $$\bar{F}_1\left(x_1\right)=\left(\frac{x_1}{\sigma_1}\right)^{-\alpha_1},\quad x_1>\sigma_1>0 ;\ \alpha_1>0.$$
	Likewise
	\[
	\bar{F}_{21}\left(x_2 \mid x_1\right)=\left(\frac{x_2}{\sigma_2\left(1+\theta u_1\right)}\right)^{-\alpha_2}, \; x_2>\sigma_2\left(1+\theta u_1\right); \; u_1=1-\left(\frac{x_1}{\sigma_1}\right)^{-\alpha_1}, \; \sigma_2>0.
	\]
	and
	$$
	\bar{F}\left(x_1, x_2\right)=\left(\frac{x_1}{\sigma_1}\right)^{-\alpha_1}\left(\frac{x_2}{\sigma_2(1+\theta(1-(\frac{x_1}{\sigma_1})^{-\alpha_1}))}\right)^{-\alpha_2}
	$$
	the survival function of a bivariate Pareto I distribution.
	
	\textbf{Case (vii):} $ \quad \alpha_i=a_i-1 ;\quad  \beta_i=-\left(a_i+1\right) ; \quad c_i=a_i b_i.$
	\par In this case,
	\[
	q_1\left(u_1\right)=\frac{b_1 a_1 u_1^{a_1-1}}{\left(1-u_1\right)^{a_1+1}} \; \Rightarrow Q_1\left(u_1\right)=\frac{b_1 u_1 ^{a_1}}{\left(1-u_1\right)^{a_1}} \; \Rightarrow \bar F_1\left(x_1\right)=\left(1+\left(\frac{x_1}{b_1}\right)^{\frac{1}{a_1}}\right)^{-1}.
	\]
	Similarly
	\[
	\bar{F}_{21}\left(x_2 \mid x_1\right)=\left(1+\left(\frac{x_2}{b_2(1+\theta u_1)}\right)^{\frac{1}{a_2}}\right)^{-1}
	\]
	and so
	\begin{equation*}
		\bar F(x_1,x_2) =\left[\left(1+(\frac{x_1}{b_1})^{\frac{1}{a_1}}\right)\left(1+(\frac{x_2}{b_2(1+\theta (1+(\frac{x_1}{b_1})^{\frac{1}{a_1}})^{-1})}\right)^{\frac{1}{a_2}}\right]^{-1},
	\end{equation*}
	which is a bivariate log logistic distribution with $a_i, b_i, x_i>0$.
	
	\par \textbf{Case (viii):} $\quad c_i=\sigma_i b_i\left(b_i+1\right),\ \alpha_i=a_i-1$ and $\beta_i=1$.
	\par Here $$q_1\left(u_1\right)=\sigma_1 b_1\left(b_1+1\right) u_1^{a_1-1 }\left(1-u_1\right),$$
	$$q_{21}\left(u_1, u_2\right)=\sigma_2 b_2\left(b_2+1\right)u_2^ {a_2-1}(1-u_2)\left(1+\theta u_1\right).$$
	Note that $q_1\left(u_1\right)$ is the quantile density function of the Govindarajulu distribution discussed in detail in \citet{nair2012govindarajulu} and accordingly $\left(q_1, q_{21}\right)$ represent a bivariate Govindarajulu distribution.
	This model does not have a tractable distribution function.
	
	All the above distributions have non-negative support. However, \citet{jones2002complementary} have identified more special univariate cases of $q_1(u)=c_1 u_1^{\alpha_1}(1-u_1)^{\beta_1}$. For example, when $\alpha_1=-\frac{1}{2},\ \beta_1=-\frac{1}{2}$ the density of $X_1$ is $f_1\left(x_1\right)=\frac{\pi}{2} \sin \pi x_1,\ 0<x_1<1$ and when $\alpha=-\frac{3}{2},\ \beta=-\frac{3}{2}$ we get a scaled $t$ distribution with 2 degrees of freedom having
	$$
	f_1\left(x_1\right)=\frac{1}{2}\left[1+\frac{x_1}{\sqrt{16 c_1{ }^2+x^2}}\right],\quad -\infty<x_1<\infty .
	$$
	Thus we can have corresponding bivariate densities as well with support as the whole $\mathbb{R}_2$ space.

	\section{ Properties of the bivariate model}\label{sec 3}
	The members of the bivariate family (\ref{2.1}) discussed in the previous section, does not appear to have been discussed in literature. Therefore, we present some important properties of these distributions that may help choice of the candidate distribution for a given set of observations. The properties of the bivariate complementary beta is discussed in some detail, since those of some other distributions can be obtained as special cases of the complementary beta. We also illustrate the case of bivariate power distribution to demonstrate as an example for the calculations in the case of tractable distribution functions.
	
	We have preferred to choose the $L$-moments to the conventional moments in describing the characteristics of the distributions in view of the advantages of the former. These are (i) the existence of the mean is sufficient for all the higher $L$-moments to exist (ii) the $L$-moments are expected values of the linear functions of order statistics that have special importance to our family (iii) they have lower sampling variance and more robustness to sampling fluctuations than the classical moments, and (iv) $L$-moment estimates are almost as good as maximum likelihood estimates in large samples and also better than the latter in small samples. For further discussions on these aspects and others we refer to \citet{hosking1990moments} and \citet{wallis1997regional}.
	
	Although the quantile functions are expressed as special functions, one can use the formulas for $L$-moments in terms of quantile densities given in \citet[p-21]{nair2013quantile} to good effect. Thus denoting the rth $L$-moment of $X_i$ by $L_{r i}, r=1,2, \ldots,\ i=1,2$ we find
	$$
	\begin{aligned}
		L_{1i} & =E\left(X_i\right)=\int_0^1\left(1-u_i\right) q_i\left(u_i\right) d u_i = \frac{\beta_i+1}{\alpha_i+\beta_i+2}\ ,  \\
		L_{2 i} & =\int_0^1\left(u_i-u_i^2\right) q_i\left(u_1\right)d u_i =\frac{\left(\alpha_i+1\right)\left(\beta_i+1\right)}{(\alpha_i+\beta_i+2)(\alpha_i+\beta_i+3)}\ ,\\
		L_{3 i} & =\int_0^1\left(3 u_i{ }^2-2 u_i^3-u_i\right) q_i\left(u_i\right) d u_i =\frac{\left(\alpha_i+1\right)\left(\beta_i+1\right)\left( \alpha_i-\beta_i\right)}{\left(\alpha_i+\beta_i+2\right)\left(\alpha_i+\beta_i+3\right)\left(\alpha_i+\beta_i+4\right)}
	\end{aligned}
	$$
	and
	\begin{eqnarray*}
		L_{4 i} & = &\int_0^1\left(u_i-6 u_i^2+10 u_i^3-5 u_i^4\right) q_i\left(u_i\right) d u_i  \\
		& = & \frac{\left(\alpha_i+1\right)\left(\beta_i+1\right)}{\left(\alpha_i+\beta_i+2\right)\left(\alpha_i+\beta_i+3\right)}\left\{1-\frac{5\left(\alpha_i+2\right)\left(\beta_i+2\right)}{\left(\alpha_i+\beta_i+4\right)\left(\alpha_i+\beta_i+5\right)}\right\}.
	\end{eqnarray*}
	Among these, $L_{1 i}$ gives the mean of $X_i$ and $L_{2 i}$, half the mean difference of $X_i$ as a measure of dispersion. The $L$-coefficient of variation is
	$$
	\tau_{2 i}=\frac{L_{2 i}}{L_{1 i}}=\frac{\alpha_i+1}{\alpha_i+\beta_i+3}<1,
	$$
	so that $X_i$ is under dispersive and $\tau_{2 i}$ decreases in $\alpha_i\left(\beta_i\right)$ for a given $\beta_i\left(\alpha_i\right)$. The $L$- skewness is
	$$
	\tau_{3 i}=\frac{L_{3 i}}{L_{2 i}}=\frac{\alpha_i-\beta_i}{\alpha_i+\beta_i+4} \text {. }
	$$
	Thus the distribution of $X_i$ is symmetric if and only if $\alpha_i=\beta_i$. Further, the positive skewness is maximum at $\beta=-2$ and negative skewness is maximum at -1 when $\alpha=-2$. Similarly the $L$-kurtosis
	$$
	\tau_{4 i}=\frac{L_{4 i}}{L_{2 i}}=\frac{\alpha_i^2+\beta_i^2-3 \alpha_i \beta_i-\alpha_i-\beta_i}{\left(\alpha_i+\beta_i+4\right)\left(\alpha_i+\beta_i+5\right)}.
	$$
	We see that the kurtosis is maximum when $\tau_{4 i} \rightarrow 1$ or $5\alpha_i \beta_i+10 \alpha_i+10 \beta_i+20 \rightarrow 0$. Otherwise it lies between $\frac{1}{4}\left(5 \tau_{3 i}^2-1\right)$ and 1.
	
	To evaluate the summary measures of the joint distribution, we use the bivariate $L$-moments and appeal to the $L$-covariance, $L$-coskewness, $L$-cokurtosis and the $L$-correlation as defined in \citet{serfling2007contribution}. In this connection from \citet{cuadras2002covariance}, we observe that if $a(x)$ and $b(x)$ are functions of bounded variation defined over the support of $X_1$ and $X_2$, and $E\left(a\left(X_1\right)\right), E\left(b\left(X_2\right)\right)$ and $E\left(a\left(X_1\right) b\left(X_2\right)\right)$ are finite then
	\begin{equation}\label{3.1}
		\operatorname{Cov}\left(a\left(X_1\right), b\left(X_2\right)\right)=\iint_{R_2}\left(\bar{F}\left(x_1, x_2\right)-\bar{F}_1\left(x_1\right) \bar{F}_2\left(x_2\right)\right) d a\left(x_1\right) d b\left(x_2\right) .
	\end{equation}
	The $L$-covariance of $X_i$ with respect to $X_j$ is
	\begin{equation}\label{3.2}
		L_{2(i, j)}=\operatorname{Cov}\left(X_i,\ 2 F_j\left(x_j\right)-1\right),\quad i, j=1,2,\ i \neq j .  
	\end{equation}
	Similarly $L$-coskewness is
		\begin{equation}\label{3.3}
			L_{3(i, j)}=\operatorname{Cov}\left(X_i,\ 6 F_j(x_j)^2-6 F_j(x_j)+1\right) 
		\end{equation}
		and the $L$-cokurtosis
		\begin{equation}\label{3.4}
			L_{4(i, j)}=\operatorname{Cov}\left(X_i,\ 20 F_j(x_j)^3-30 F_j(x_j)^2+12 F_j(x_j)-1\right).  
	\end{equation}
	In the sequel, we calculate the measures of $X_1$ with respect to $X_2$ only as the other follows similarly. Recall that, in general, the quantile functions involved are
	$$
	Q_1\left(u_1\right)=I_{u_1}\left(\alpha_1+1, \beta_1+1\right) \text { and } Q_{21}\left(u_1, u_2\right)=\frac{{I}_{u_2}\left(\alpha_2+1, \beta_2+1\right)}{1+\theta u_1}.
	$$
	The $L$-covariance of $X_1$ with respect to $X_2$ is from (\ref{3.1}) and (\ref{3.2}),
	$$
	L_{2(1,2)}=2 \iint_R\left(\bar{F}_1\left(x_1\right) \bar{F}_{21}\left(x_2 \mid x_1\right)-\bar{F}_1\left(x_1\right) \bar{F}_1\left(x_2\right)\right) f_2\left(x_2\right) d x_1 d x_2.
	$$
	Applying the transformations $x_1=Q_1 (u_1)$ and $x_2=Q_2 (u_2)$,
	\begin{equation}\label{3.5}
		L_{2(1,2)}=2 \int_0^1 \int_0^1\left[\left(1-u_{21}\right)\left(1-u_1\right)-\left(1-u_1\right)\left(1-u_2\right)\right]q_1(u_1) d u_1 d u_2.   
	\end{equation}
	Similarly from (\ref{3.3}) and (\ref{3.4}),
	\begin{equation}\label{3.6}
		L_{3(1,2)}=\int_0^1 \int_0^1\left[\left(1-u_{21}\right)\left(1-u_1\right)-\left(1-u_1\right)\left(1-u_2\right)\right]\left(12 u_2-6\right)q_1(u_1) d u_1 d u_2   
	\end{equation}
	and
	\begin{equation}\label{3.7}
		L_{4(1,2)}=\int_0^1 \int_0^1\left(\left(1-u_{21}\right)\left(1-u_1\right)-\left(1-u_1\right)\left(1-u_2\right)\right)\left(60 u_2^2-60 u_2+12\right)q_1(u_1) d u_1 d u_2,  
	\end{equation}
	where $ u_{21}=F_{21}\left(Q_{21}\left(u_1, u_2\right)\right)=F_{21}\left(Q_2 \mid X_1>Q_1\right) =x_2 \frac{I^{-1} _{u_2}\left(\alpha_2+1, \beta_2+1\right)}{1+\theta u_1} $.
	\par The above expressions have to be evaluated numerically for the bivariate beta complementary model. However, when $F_{21}$ has a tractable form the evaluation is done as in the case of given below for the bivariate power distribution, shown below. Here,
	$$
	F_{21}\left(x_2 \mid x_1\right)=\left(\frac{x_2}{b_2(1+\theta u_1)}\right)^{a_2},
	$$
	so that
	$$
	u_{21}=\frac{u_2}{\left(1+\theta u_1\right)^{a_2}}\ .
	$$
	Hence
	$$
	\begin{aligned}
		L_{2(1,2)} & =2 \int_0^1\int_0^1\left[\left(1-u_1\right)\left(1-\frac{u_2}{\left(1+\theta u_1\right)^{a_2}}\right)-\left(1-u_1\right)\left(1-u_2\right)\right]q_1(u_1) d u_1 d u_2 \\
		& =-c_1 a_1\,_2F_1\left(\frac{1}{a_1},a_2,1+\frac{1}{a_1},-\theta \right) - \frac {_2F_1\left(1+\frac{1}{a_1},a_2,2+\frac{1}{a_1},-\theta \right)+a_1}{1+a_1}.
	\end{aligned}
	$$
	The other two measures $L_{3(1,2)}$ and $L_{4(1,2)}$ can be obtained in a similar manner. It may be of interest to note that coskewness and cokurtosis measures the extent to which the random variables $X_1$ and $X_2$ change together with positive (negative) skewness referring to variables having positive (negative) deviations at the same time. Some areas in which they are extensively used are ranking portfolios, asset pricing and hydrology. See for example \citet{christie2001coskewness} and \citet{zsolt2021co} and their references.\\
	Finally, the $L$-correlation of $X_1$ towards $X_2$ is
	$$
	\begin{aligned}
		\rho_{12} & =\frac{L_{2(1,2)}}{L_{21}} \\
		& =-\frac{(6+5\alpha_1+\alpha_1^2)}{(\alpha+1)}\,_2F_1\left(\alpha_1+1,\frac{1}{\alpha_2+1},\alpha_1+2,-\theta \right)\\
		& \quad  - \frac {_2F_1\left(\alpha_1+1,\frac{1}{\alpha_2+1},\alpha_1+3,-\theta \right)(\alpha_1+1)+1}{(\alpha_1+1)}.
	\end{aligned}
	$$
	The $L$ correlation $\rho_{21}$ of $X_2$ towards $X_1$, is in general not the same as $\rho_{12}$. Both lies in $[-1,1]$ with the extreme values of $\rho_{12}$ is attained when $X_1$ is an increasing (decreasing) function of $X_2$.
		
	\section{Application to real data}\label{sec 4}
	
	To demonstrate the utility of our distribution (\ref{2.1}) in real life situations, we show that it explains satisfactorily the data generating mechanism for two data sets from distinct contexts. The first one originates from life tests of two types of cable installation reported in \citet[p-264]{lawless2011statistical}. Of the specimens of each type tested, the last observation in each being a censoring time, they  removed this leaves the observations in Table \ref{table1}.
	\begin{table}[ht]
		\centering
		\begin{tabular}{|l|l|l|l|l|l|l|l|l|l|}
			\hline
			$X_1$ & 5.1 & 9.2  & 9.3  & 11.8 & 17.7 & 19.4 & 22.1 & 26.7 & 37.3 \\ \hline
			$X_2$ & 11  & 15.1 & 18.3 & 24   & 29.1 & 38.6 & 44.2 & 45.1 & 50.9 \\ \hline
		\end{tabular}
		\caption{Lifetime of cable installation}
		\label{table1}
	\end{table}
	For reasons mentioned in Section \ref{sec 3}, We have computed the parameters of the model using the method of $L$-moments. The $L$-moments approach involves calculating the parameters from the equations  $L_i = l_i, i = 1, 2,...$, where $l_i$ is the $i^{th}$ sample $L$-moment which has the formula $$
	l_i=\frac{1}{n} \sum_{j=0}^{r-1} \frac{(-1)^{1-i-j}(i+j-1) !}{(j !)^2(i-j-1) !}\left(\sum_{r=1}^n \frac{(r-1)(j)}{(n-1)(j)}\right), i=1,2,3...
	$$
	To estimate the parameters, the first three sample $L$-moments are compared with the corresponding population $L$-moments.
	The first three $L$-moments of $X_i$'s are respectively given by
	$$\begin{aligned}
		& L_{1 i}=c_i\frac{\Gamma(\alpha_i+1)\Gamma(\beta_i+2)}{\Gamma(\alpha_i+\beta_i+3)},\\
		& L_{2 i}=L_{1 i}=c_i(\beta_i+2)\frac{\Gamma(\alpha_i+2)\Gamma(\beta_i+1)}{\Gamma(\alpha_i+\beta_i+4)},\\
		& L_{3 i}=c_i(\alpha_i-\beta_i)\frac{\Gamma(\alpha_i+2)\Gamma(\beta_i+2)}{\Gamma(\alpha_i+\beta_i+5)}.\\
	\end{aligned}$$ 	
	Assuming that we have $ n $ copies $X_i$ namely, $\left(X_{i1} ,X_{i2} , \ldots X_{i n}\right), i=1,2,$ the estimates of the parameters of the marginals are obtained as
	$$
	\begin{aligned}
		& \hat{c_1}=9.0819,\ \ \hat{\alpha}_1=0.4864,\ \ \hat{\beta}_1=0.9946 \\
		& \hat{c}_2=29.2295,\ \hat{\alpha}_2=0.3406,\ \text { and } \hat{\beta}_2=0.3531 .
	\end{aligned}
	$$
	To estimate the parameter $\theta$, we have equated the product moment $E\left(X_1 X_2\right)$ with its sample counterpart and found $\hat{\theta}=0.6821$. Accordingly, the bivariate distribution specified in \eqref{2.1} becomes
	\begin{equation}\label{4.1}
		\begin{aligned}
			q_1(u_1)&=9.0819 u_1^{0.4864}(1-u_1)^{0.9946}, \; \text{ and }\\
			q_{21}(u_1, u_2)&=(1+0.68210 u_1)29.2295u_2^{0.3406}(1-u_2)^{0.3531}.
		\end{aligned}
	\end{equation}
	
\begin{figure}[ht]
		\centering
		\includegraphics[width=8cm]{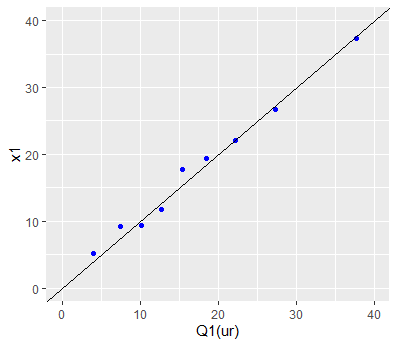}
		\caption{Q-Q plot for the distribution of $X_1$}
		\label{fig1}\end{figure}

	In order to test the goodness-of-fit we have used two methods, the $Q-Q$ plot as a visual aid and the Kolmogorov-Smirnov (K-S) test. The $Q-Q$ plot for the distribution of $X_1$, given in Figure \ref{fig1}, indicate that the sample plots cluster around the bisector to justify that the $Q_1\left(u_1\right)$ represents the data satisfactorily. The K-S test statistic
	$$
	D_1=\sup _{x_1}\left|F_n\left(x_1\right)-F_1(x)\right|,
	$$
	where $F_n\left(x_1\right)$ is the empirical distribution function of $X_1$ calculated from the sample $\left(X_{11}, X_{12} \cdots, X_{1 n}\right)$. Since $F_1\left(x_1\right)$ is not available explicitly, we solve the equation $\hat{Q}_1\left(u_{1 i}\right)=x_{1 i}$ to find $u_{1 i}=F_1\left(x_{1i}\right)$, using $\hat{Q}_1$ as the estimate of $Q_1\left(u_1\right)$ when the parameters $c_1, \alpha_1$ and $\beta_1$ are replaced by their estimates. This gives $D_1=0.097$, indicating that the fit is satisfactory.
	
\begin{figure}[ht]
		\centering
		\includegraphics[width=8cm]{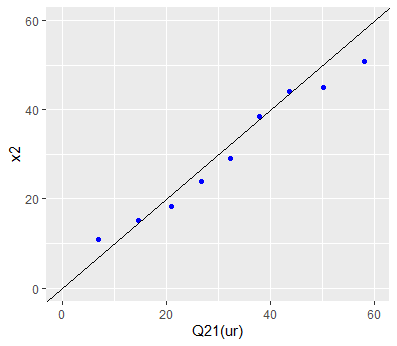}
		\caption{Q-Q plot for the distribution of $X_2\mid X_1>x_1$}
		\label{fig2}\end{figure}

	The same tests were employed fer testing the distribution of $X_2$ given $X_1>x_1$ represented by $q_{21}$. From the estimated values of $u_{1 i}$ obtained in the first case we solve $\hat{Q}_{21}\left(u_{2 i}\right)=x_{2 i}$ for $u_{2 i}$ which is the representative of $F_{21}\left(x_{2 i}\mid x_{1}\right)$. For each sample value $x_{1 i}$ we compute the K-S statistic
	$$
	D_{21, i}=\sup | F_n\left(x_{2} \mid x_{1 i}\right)-F_{21}\left(x_2 \mid x_{1 i}\right)|,
	$$
	to verify the goodness-of-fit. A typical case when $x_{i i}$ is the smallest in the sample gives $D_{21,1}=0.155$, so that the assumption of the distribution is not rejected. The $Q-Q$ plot given in Figure \ref{fig2} also  supports the same evidence. Thus the $Q-Q$ plots and K-S tests does not reject the bivariate complementary beta distribution for the given data.
	The values of mean, standard deviation, $L$-covariance, $L$-correlation, $L$-coskewness and $L$-cokurtosis for the data were obtained as 
	\begin{equation*}
		\begin{aligned}
			&E(X_1)= 17.62,\quad \text{Standard Deviation}(X_1)= 10.14\\
			&E(X_2)=30.70 ,\quad \text{Standard Deviation}(X_2)= 14.54, \\
			&L\text{-covariance}= 0.24, \quad L\text{-correlation}= 0.53 , \\
			&L\text{-coskewness}= 0.30 \text{ and }   L\text{-cokurtosis} = 0.25.
		\end{aligned}   
	\end{equation*}
	Since the variance of $X_1$ and $X_2$ are relatively larger than the respective mean, we conclude that the variables are likely to be overdispersive. The $L$-covariance is positive indicates that the variables move in the same direction but they have only a moderate relationship as indicated by the $L$-correlation of 0.53. The $L$-coskewness is positive is indicative of the fact that the variables undergo positive deviation at the same time. On the otherhand the $L$-cokurtosis being small shows that there are no extreme positive or negative deviations at the same time.
	
	The second dataset was provided by \citet{kim2004reliability}, who tracked the failure times of 20 sample units from a three-component system.  Since the present study considers the modelling of bivariate data, we have taken only the failure times of two components for the analysis as given in Table \ref{table2}.
	\begin{table}[ht]\scalebox{0.65}
		{\begin{tabular}{|l|l|l|l|l|l|l|l|l|l|l|l|l|l|l|l|l|l|l|l|l|}
				\hline
				$X_1$  &  0.37 &  0.06 &  0.2 &  1.62 &  5.7  &  2.25 &  2.5  &  2.44 &  0.12 &  0.79 &  7.22 &  2.81 &  4.13 &  5.67 &  0.96 &  7.16 &  0.32 &  7.32 &  2.58 &  1.73 \\ \hline
				$X_2$
				&  6.93 &  2.42 &  0.2 &  2.34 &  1.96 &  4.6  &  0.09 &  7.27 &  0.06 &  8.61 &  1.38 &  5.05 &  0.52 &  1.11 &  3.54 &  2.38 &  1.89 &  1.54 &  8.61 &  1.22\\ \hline
		\end{tabular}}
		\caption{Failure times two-components as reported in \citet{kim2004reliability}}
		\label{table2}
	\end{table}
	
\begin{figure}[ht]
		\centering
		\includegraphics[width=8cm]{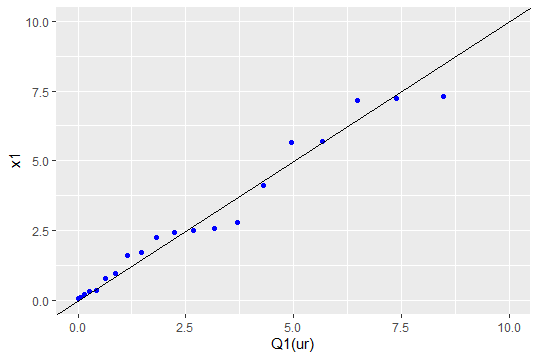}
		\caption{Q-Q plot for the distribution of $X_1$}
		\label{fig3}\end{figure}

\begin{figure}[ht]
		\centering
		\includegraphics[width=8cm]{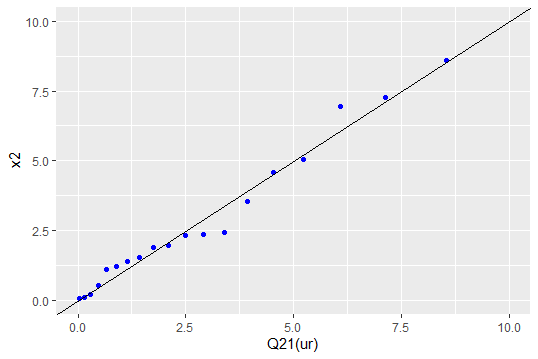}
		\caption{Q-Q plot for the distribution of $X_2\mid X_1>x_1$}
		\label{fig4}\end{figure}

	We calculate the estimates of the parameters as in the first case, to obtain 
	$$
	\begin{aligned}
		& \hat{c}_1=13.0499,\ \hat{\alpha}_1=0.8856,\ \hat{\beta}_1=-0.1844 \\
		&\hat{c}_2=5.9257,\ \hat{\alpha}_2=0.3555,\ \hat{\beta}_2=-0.6695 \text{ and } \hat{\theta}=0.5492.
	\end{aligned}
	$$
	The Q-Q plots corresponding to $Q_1$ and $Q_{21}$ are exhibited in Figures \ref{fig3} and \ref{fig4} . Further, the K-S test values are $D_1=0.110$ and $D_{21}=0.133$ with respective $p$-values 0.310 and 0.277. Thus the data does not reject the bivariate complementary beta distribution with 
	\begin{equation}\label{4.2}
		\begin{aligned}
			q_1(u_1)&=13.0499 u_1^{0.8856}(1-u_1)^{-0.1844} \; \text{ and } \\
			q_{21}(u_1, u_2)&=(1+0.5492 u_1)5.9257u_2^{0.3555}(1-u_2)^{-0.6695}.
		\end{aligned}
	\end{equation} 
	Thus the second dataset also fits the model given in \eqref{2.1}.  
	
	Using models (\ref{4.1}) and (\ref{4.2}) one can easily derive the desired characteristics of the distribution of $(X_1,X_2)$ based on the expressions in Section \ref{sec 3} and the probabilities of events of interest. 
	
\begin{figure}[ht]
		\centering
		\includegraphics[width=8cm]{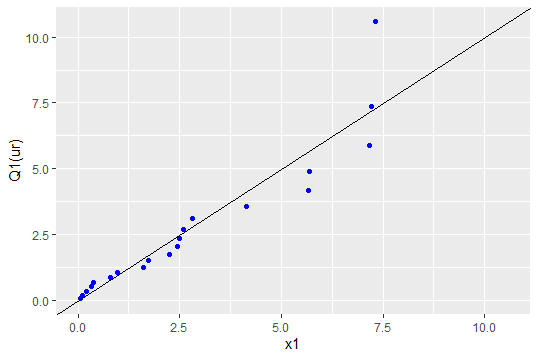}
		\caption{Q-Q plot for the distribution of $X_1$}
		\label{fig5}\end{figure}

\begin{figure}[ht]
		\centering
		\includegraphics[width=8cm]{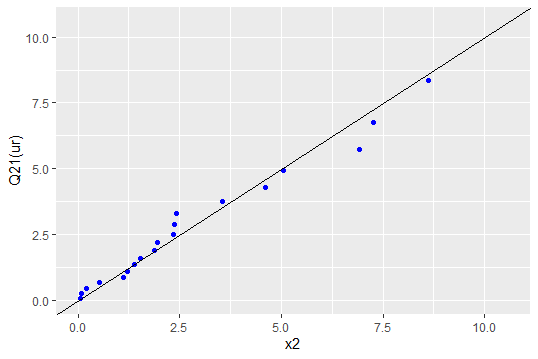}
		\caption{Q-Q plot for the distribution of $X_2\mid X_1>x_1$}
		\label{fig6}\end{figure}

	To examine the usefulness of the proposed model compared to some existing bivariate models, we proceed to fit the same data to the  bivariate linear mean residual quantile model defined by \cite{vineshkumar2019bivariate}, given by
		\begin{equation}\label{4.3}
			Q_1\left(u_1\right) = -\left(a_1+b_1\right) \log \left(1-u_1\right)-2 b_1 u_1, 0 \leq u_1 \leq 1, \, a_1>0,\, \left(a_1+b_1\right)>0
		\end{equation}
		and
		\begin{eqnarray}\label{4.4}
			Q_{21}\left(u_2 \mid u_1\right) = - &\left(a_2+c+\left(b_2+d\right) u_1\right) \log \left(1-u_2\right)-2\left(c+d u_1\right) u_2,\nonumber\\
			& 0 \leq u_2 \leq 1,\, a_2+c>0,\, a_2>0,\, a_2+b_2 \geq c+d.
		\end{eqnarray}
		The above bivariate model also do not have a closed form distribution functions.  The method of $L$-moments are again employed for the estimation of parameters of the model, which are obtained as 
		\[
		\hat{a_1}=2.798, \, \hat{b_1}=0.159, \,\hat{a_2}=3.086, \, \hat{c}=0.086, \hat{b_2}=4.628, \, \text{ and }\hat{d}=-7.16,
		\]
		Then the quantile functions \eqref{4.3} and \eqref{4.4} become
		\begin{equation}\label{4.5}
			Q_1\left(u_1\right) = -\left(2.798+0.159\right) \log \left(1-u_1\right)-2* 0.159 *u_1 
		\end{equation}
		and
		\begin{equation}\label{4.6}
			Q_{21}\left(u_2 \mid u_1\right)=-\left(3.086+0.086+\left(4.628-7.16\right) u_1\right) \log \left(1-u_2\right)-2\left(0.086-7.16* u_1\right) u_2.
		\end{equation}
		The $Q-Q$ plots corresponding to $Q_1$ and $Q_{21}$ in \eqref{4.5} and \eqref{4.6} are given in Figures \ref{fig5} and \ref{fig6} and the corresponding K-S test statistics values are $D_1=0.126$ and $D_{21}=0.322$ respectively. Thus the data does not reject the bivariate linear mean residual quantile function. Clearly, the K-S statistic value for the bivariate complementary beta model specified in \eqref{4.2} is small compared to the bivariate mean residual quantile model.  This allow us to conclude that the proposed bivariate model is a better model for modelling the given bivariate data set.
	\newpage
	\section{Concluding Remarks}\label{sec 5}
	We have presented a flexible family of bivariate distribution based on a bivariate quantile function.  The proposed family of bivariate distribution is characterized by the use of quantile functions. The model employs two primary functions: $Q_1\left(u_1\right)$, which represents the quantile function of the first variable $X_1$, and $Q_{21}\left(u_1, u_2\right)$, which is the quantile function of the second variable $X_2$ given $Q_1(u_1)$. This framework allows for the derivation of the marginal distribution $F_1\left(x_1\right)$ and the conditional distribution $F_2\left(x_2 \mid x_1\right)$, leading to the joint distribution $F\left(x_1, x_2\right)$. The model is highly flexible, allowing for the representation of a wide range of bivariate distributions by simply altering the parameter values. This versatility enables the modeling of various dependency structures and tail behaviors, making it applicable to numerous fields. The adaptability of the model to different bivariate distributions without requiring closed form expressions makes it suitable for a wide range of practical applications, providing a unified framework for bivariate analysis and makes the model more straightforward to implement and interpret.  As quantile functions are less sensitive to outliers and extreme values compared to traditional distribution functions, enhances the model's reliability and accuracy, particularly in heavy-tailed distributions and datasets with anomalies. 
	
	Generating random samples from the proposed bivariate distribution is efficient and straightforward. By simulating uniform random variables $u_1$ and $u_2$, we can easily obtain samples of $X_1$ and $X_2$ through the quantile functions $Q_1\left(u_1\right)$ and $Q_{21}\left(u_1, u_2\right)$. This property is particularly beneficial in scenarios requiring large-scale simulations, where the computational efficiency and simplicity of the quantile-based approach significantly reduce the complexity.  The use of quantile functions allows for the derivation of analytical expressions for marginal and conditional distributions, enhancing the model's tractability. This facilitates the calculation of probabilities, moments, and other statistical measures, which are essential for practical applications and theoretical studies and simplifies the analysis and understanding of the dependence structure between $X_1$ and $X_2$.  
	
	Additionally, our model facilitates the use of $L$-moments for parameter estimation, offering several advantages. $L$-moments, which are linear combinations of order statistics, provide more robust estimates compared to traditional moments, especially in the presence of outliers or small sample sizes. One of the significant properties of $L$-moments is that if the mean of a distribution exists, then all higher-order $L$-moments also exist. This ensures that $L$-moments can be reliably used for a wide range of distributions. Their use in our quantile-based model enhances the estimation process, making it more robust and reliable.  Several additional properties, characterizations and applications of the new distribution require separate consideration, in other areas require further examination.
	\section*{Acknowledgements}
	The first author wish to thank the Cochin University of Science and Technology,
	India for carrying out this research work.
	
	\section*{Conflict of interest statement}
	On behalf of all the authors, the corresponding author states that there is no conflict of interest.
	\bibliographystyle{apalike}
	\bibliography{myref}
\end{document}